\documentclass[]{interact}

\usepackage{epstopdf}
\usepackage[caption=false]{subfig}

\usepackage[numbers,sort&compress]{natbib}
\bibpunct[, ]{[}{]}{,}{n}{,}{,}
\renewcommand\bibfont{\fontsize{10}{12}\selectfont}

\usepackage{algorithm}
\usepackage{algpseudocode}

\theoremstyle{plain}

\theoremstyle{definition}

\theoremstyle{remark}

\usepackage{multicol}
\usepackage{hyperref}
\begin{document}


\title{The Relationship between Moderate to Vigorous Physical Activity and Metabolic Syndrome: A Bayesian Measurement Error Approach  \footnote{This is an Accepted Manuscript of an article published by Taylor \& Francis in Journal of Applied Statistics on May 12, 2022, available at: \url{https://doi.org/10.1080/02664763.2022.2073336}} }

\author{
\name{Daniel Ries\textsuperscript{a}\thanks{CONTACT D.~Ries. Email: dries@sandia.gov} \footnote{} and Alicia Carriquiry\textsuperscript{b}}
\affil{\textsuperscript{a}Statistics and Data Analytics Department, Sandia National Laboratories, Albuquerque, NM} 
\affil{\textsuperscript{b}Department of Statistics, Iowa State University, Ames, IA} 
}

\maketitle

\begin{abstract}
Metabolic Syndrome (MetS) is a serious condition that can be an early warning sign of heart disease and Type 2 diabetes. MetS is characterized by having elevated levels of blood pressure, cholesterol, waist circumference, and fasting glucose. There are many articles in the literature exploring the relationship between physical activity and MetS, but most do not consider the measurement error in the physical activity measurements nor the correlations among the MetS risk factors. Furthermore, previous work has generally treated MetS as binary, rather than directly modeling the risk factors on their measured, continuous space. Using data from the National Health and Nutrition Examination Survey (NHANES), we explore the relationship between minutes of moderate to vigorous physical activity (MVPA) and MetS risk factors. We construct a measurement error model for the accelerometry data, and  then model its relationship between MetS risk factors with nonlinear seemingly unrelated regressions, incorporating dependence among MetS risk factors. The novel features of this model give the medical research community a new way to understand relationships between MVPA and MetS.  The results of this approach present the field with a different modeling perspective than previously taken and suggest future avenues of scientific discovery. 

\end{abstract}


\section{Introduction}

Epidemiologists and health care professionals, among others, are interested in the relationship between physical activity and metabolic syndrome. Metabolic syndrome (MetS) is a group of risk factors that increase the chances of diseases such as cardiovascular disease and Type 2 diabetes \cite{kim2}. According to the National Cholesterol Education Program Adult Treatment Panel, the risk factors that comprise MetS include abdominal obesity (waist circumference $>$102 cm in men and $>$88 cm in women), high triglyceride level ($>$150 mg/dL), high fasting glucose level ($>$110 mg/dL), high systolic ($>$130 mm Hg) or high diastolic blood pressure ($>$85 mm Hg), and  low levels of  high-density lipoprotein (HDL) ($<$40 mg/dL for men, $<$50 mg/dL for women) \cite{grundy2005}. A person is said to have MetS when he or she exhibits three or more elevated risk factors as described above. However, this definition has changed and varies slightly depending on health agency \cite{kassi2011}. Although not directly included in the definition of MetS, high levels of low-density lipoprotein (LDL) cholesterol ($>$ 160 mg/dL) are also considered a health risk \cite{sisson, tucker}.   MetS is a serious problem in the United States as it affects 34\% of US adults \cite{moore}.

There have been numerous studies examining the relationship between physical activity and MetS. Franks et al. \cite{franks} adjusted responses for measurement error and used multiple  and logistic regression and found a moderate, negative association between level of physical fitness and MetS. Ford et al. \cite{ford} and Dalacorte et al. \cite{dalacorte} both calculated odds ratios for MetS adjusted for covariates such as age and sex for prespecified levels of physical activity. 
Ford et al. \cite{ford} focused on U.S. adults and reached similar conclusions as previously mentioned papers. Dalacorte et al. \cite{dalacorte} considered Brazilian elders and found no associations between physical activity levels and MetS. Sisson et al. \cite{sisson} explored the relationship between daily steps and odds of having MetS, and found that the odds of having MetS were 10\% lower for each additional 1000 steps. Camhi et al. \cite{camhi} looked at the relationship between moderate intensity lifestyle activity (lower than moderate to vigorous physical activity (MVPA)) and MetS and  found that more time spent at this activity level, independent of MVPA, is associated with lower odds of MetS. Tucker et al. \cite{tucker} examined the relationship between self-reported and accelerometer-measured physical activity and MetS. Using logistic regression, they found that individuals who do not meet the \emph{Physical Activity Guidelines for Americans} have greater odds of having MetS, (odds ratio of 2.57 for men and 4.40 for women when using accelerometer data.) Huang et al.  \cite{huang} examined the odds ratios of MetS and low/high levels of leisure, occupational, and commuting physical activity in Taiwanese workers and found leisure and occupational physical activity to be associated with MetS. Tucker et al.  \cite{tucker} and Huang et al. \cite{huang} were the only papers that looked at odds ratios of individual MetS risk factors.

Most of these studies considered the relationship between the binary diagnosis of MetS and physical activity levels for  individuals, typically using cut-points for what is considered ``high levels" of risk factors. Because the problem is  defined as binary, some form of logistic regression using  physical activity levels from accelerometers or self-reports are typically used to calculate odds ratios for different subpopulations. We propose to model these risk factors on the continuous scale as they are collected, rather than on the typical ``high" or ``normal" levels they are often attributed.  

The main contribution of this paper is a novel modeling approach exploring the relationship between MVPA and all the MetS risk factors  while accounting for dependencies and measurement errors often ignored. We introduce a model that allows for non-linear relationships between MVPA and each risk factor and accounts for dependence across  risk factors for individuals via seemingly unrelated regressions (SUR) \cite{zellner}. Many analyses on MetS assume independence across risk factors and fail to capture critical dependencies. Unlike most of the studies we have identified, we construct a measurement error model (MEM) for the physical activity observations before we assess the relationship between time in MVPA and MetS risk factors in a risk factor model (RFM), since physical activity data is often collected via accelerometers or self-report, both of which exhibit measurement error. 
All of these model features are tied together in a Bayesian model to provide easy uncertainty quantification and to provide simple interpretations.

This paper is structured as follows: Section 2 introduces the NHANES data and adjustments for nuisance effects and the handling of missing data. Section 3 shows the MEM for the physical activity data and the MetS RFM relating physical activity to MetS risk factors. Section 4 presents the results and model validation. Section 5 discusses the results and proposes future research directions.

\section{Data}

\subsection{National Health and Nutrition Examination Survey}

\begin{figure}[h]
\centering
\includegraphics[width=.7\textwidth]{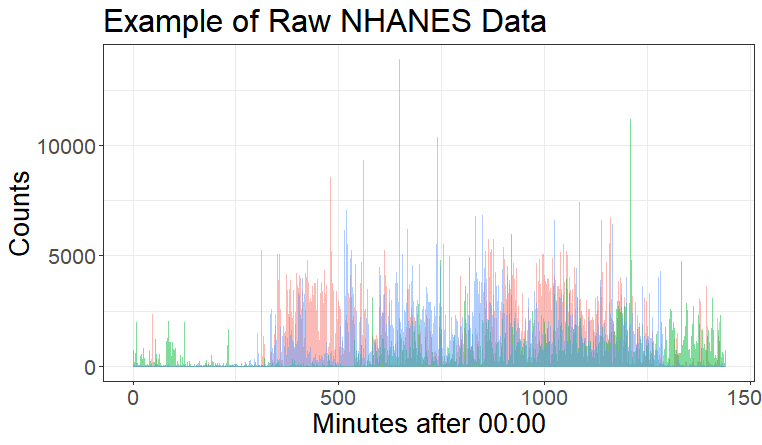}
\caption{Raw NHANES accelerometry data for three individuals (blue,red,green curves) for a single day. Raw response is ``counts." }
\label{raw}
\end{figure}

\begin{figure}[h]
\centering
\includegraphics[width=.7\textwidth]{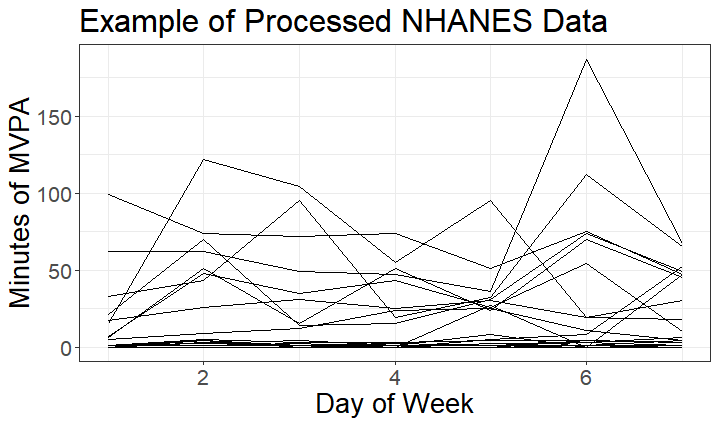}
\caption{Processed NHANES accelerometry data to MVPA for 20 individuals.  }
\label{processed}
\end{figure}

The National Health and Nutrition Examination Survey (NHANES)  is a large, national survey that is used to assess the health of adults and children in the United States. The last survey with accelerometry measurements  is 2003-2006. Using  data from accelerometers rather than self-report gives a more objective measure of activity. NHANES \cite{nhanes} participants were instructed to wear the ActiGraph AM-7164 for seven straight days during waking hours and to remove the device only while sleeping or during any water activities. The NHANES website states that the device records uniaxial movements, which means that activities including stationary bikes, ellipticals, or primarily upper body movements may not be recorded accurately. The accelerometer records measurements of the physical intensity  by minute. The intensity of the activity is reported as  \emph{counts} where more counts correspond to more intense activity, and zero counts correspond to no activity. Figure \ref{raw} shows  an example of raw accelerometry data for three different individuals for a single day.  If the number of counts in a minute exceeds 2020, then the individual is said to be engaged in moderate physical activity; if the number of counts exceeds 5999, the activity is said to be vigorous \cite{tucker}, and if the number of counts is zero, the individual has not participated in physical activity. We calculate daily minutes in MVPA for each individual using these thresholds.  Figure \ref{processed} shows processed accelerometry data to daily minutes in MVPA for 20 individuals.

In this study we consider adults age 18+. There are 2508 individuals with accelerometry data on all 7 days of observation. Analyses in the past using NHANES accelerometer data typically follow a general rule, like a ``valid" day is considered as having at least 10 hours of wear time \cite{sisson, tucker}. We follow this rule to denote a ``full" day. Table \ref{missing} shows demographic summaries for the 2508 NHANES participants with full accelerometry data. All of these demographic variables will be included as covariates in the MEM.

\begin{table}[h]
\centering
\begin{tabular}{l|cc}
\hline
   & Men & Women \\
\hline
{\bf Age} & & \\
18-34 & 198 & 203 \\
35-49 & 274 & 270\\
50-65 & 333 & 366\\
66-85 & 460 & 404\\
\hline
{\bf Race/Ethnicity}& &\\
Mexican American & 236 & 209 \\
Other Hispanic & 22 & 25 \\
Non-Hispanic White & 764 & 751 \\
Non-Hispanic Black & 192 & 199 \\
Other Race & 51 & 59 \\
\hline
{\bf Education} \\
$<9^{th}$ Grade &193 & 137\\
$>9^{th}$ Grade (No diploma) &128 &135 \\
HS Diploma/GED &294 &322 \\
Some college or AA degree &325 & 358\\
College graduate & 323& 291\\
Other & 2 & 0 \\
\hline
{\bf BMI} & & \\
Mean &27.5 & 28.0\\
SD & 4.7& 6.2\\
\hline
\end{tabular}
\caption{Demographic summaries for the NHANES sample with complete MetS risk factor and accelerometery data.}
\label{missing}
\end{table}

As part of this health assessment, NHANES also measures each MetS risk factor for a subset of the study's sample. There are 1039 individuals who have full accelerometer data as well as full MetS risk factor data (waist circumference, HDL,  systolic and diastolic blood pressure, glucose, triglycerides, and LDL).  These 1039 individuals will be used for the RFM (note the MEM has a larger sample size, but the RFM sample is inclusive to the MEM sample, allowing us to estimate the MEM as best as possible). Table \ref{mets} gives summaries of the 1039 individuals with full MetS risk factor and accelerometry data by sex. 

\begin{table}
\centering
\begin{tabular}{l|cc}
\hline
 & Men & Women \\
\hline
Waist Circumference (cm) & 98 (15) & 95 (15) \\
Glucose (mg/dL) & 104 (29) & 100 (30) \\
Triglycerides (mg/dL) & 134 (71)  & 127 (69) \\
Systolic Blood Pressure (mm Hg) & 125 (17) & 121 (21) \\
Diastolic Blood Pressure (mm Hg) & 69 (12) & 67 (12) \\
LDL (mg/dL) & 115 (35) & 114 (36)  \\
HDL (mg/dL) & 50 (13) & 60 (16) \\
\hline
\end{tabular}
\caption{MetS risk factor mean values (sd) for the NHANES sample with complete MetS risk factor and accelerometry data.}
\label{mets}
\end{table}

\begin{figure}
\centering
\includegraphics[width=\textwidth]{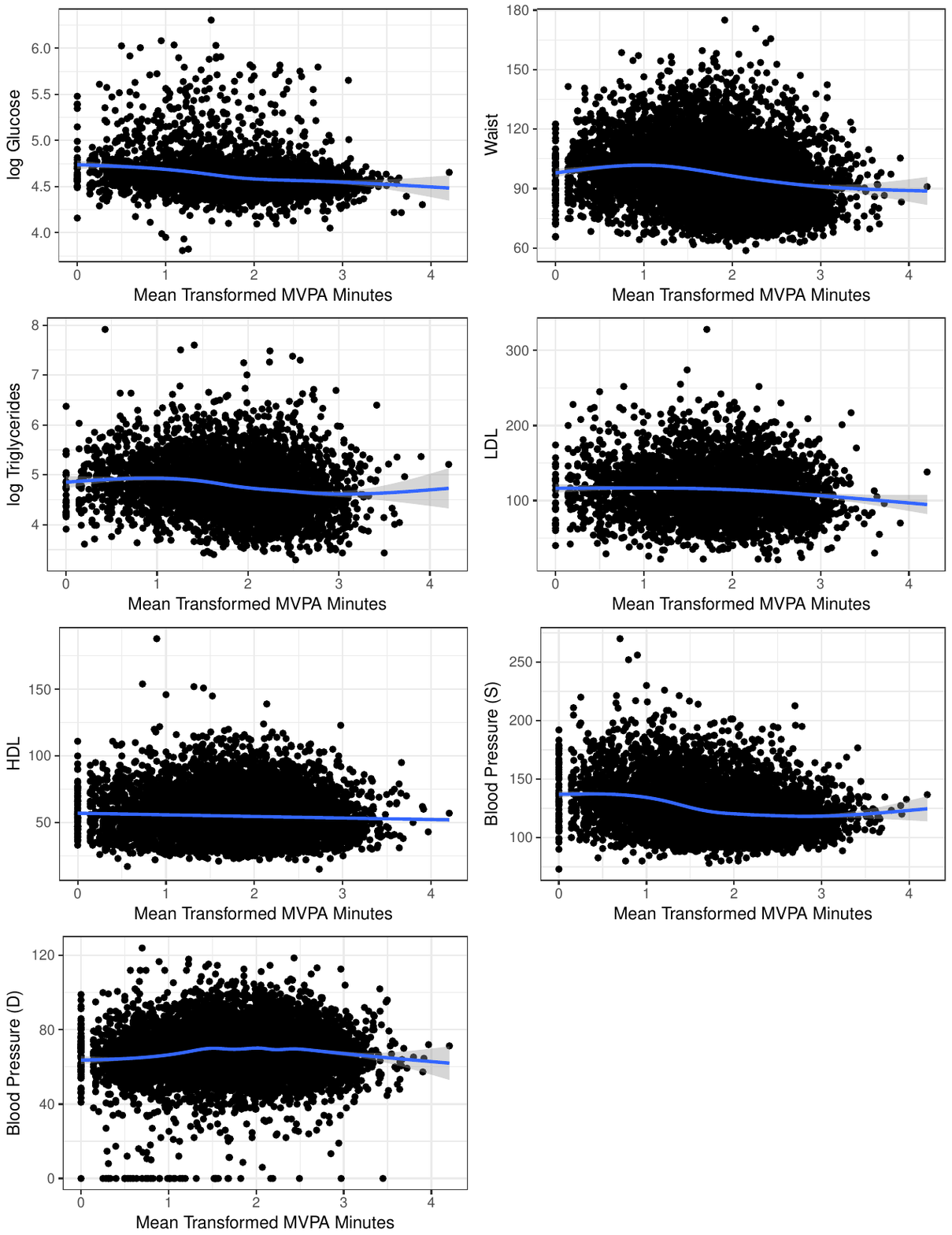}
\caption{Observed mean fourth-root minutes in MVPA against MetS risk factors. Loess curve drawn on each plot to see trend. }
\label{scatter}
\end{figure}

Figure \ref{scatter} shows the mean of fourth root of minutes in MVPA for each individual versus each risk factor. The fourth-root transformation helps satisfy the MEM assumptions described in Section \ref{sec:assumptions}. A loess curve on top of each plot to gives an understanding of the relationships. Log transformations help the heavily skewed Glucose and Triglycerides measurements. Although there is large between-individual variability, there are relationships with practical changes in these data at the population level.

\subsection{Adjustments for Weekend Effects}
\label{sec:dayadjustments}

Let values 1-7 identify each of the seven days of the week (M-Su) during which individuals wore the accelerometer. Let $W_{0ij} \in [0,\infty)$ represent the minutes in MVPA for participant $i$ on day $j$.   In this study we are interested in the long term relationship between MVPA and MetS risk factors, not time in MVPA during any given day, therefore we account for weekend effects on MVPA. Since we saw differences between Saturday and Sunday, we adjust for those effects individually. There were no significant differences between average MVPA levels as reported by accelerometry according to an ANOVA. 

We define a new variable,  $W_{1ij} \in [0,\infty)$ representing MVPA on day $j$ adjusted for Saturday and Sunday effects and estimate it using a linear model where observed minutes in MVPA is regressed on  an indicator for activity on Saturday and an indicator for activity on Sunday.  The adjustment  is a ratio, rather than  linear  to avoid negative adjusted MVPA. This follows closely from \cite{nusser}. The adjustment model is:

\begin{align}
W_{0ij} &= \psi_0 + Saturday_{ij} \psi_1 + Sunday_{ij} \psi_2  + \kappa_{ij}, \\
\kappa_{ij} &\sim N(0,\sigma_{\kappa}^2), \\
\hat{W}_{0ij} &= \hat{\psi}_0 + Saturday_{ij}\hat{\psi}_1 + Saturday_{ij}\hat{\psi}_2, \\
W_{1ij} &= \hat{W}_{0ij}^{-1} \bar{W}_{0 \cdot \cdot} W_{0ij},
\end{align}

where  $\bar{W}_{0 \cdot \cdot}$ is the mean of the observed minutes in MVPA and $Saturday_{ij}$ and $Sunday_{ij}$ are indicator variables for whether day $j$ was Saturday or Sunday for individual $i$, respectively. The main difference from the adjustment made in \cite{nusser} is here we adjust to the overall mean because we are interested usual MVPA across all days, not just weekdays. 

\subsection{Adjustments for Survey Weights}

NHANES data are collected using an unequal probability survey design, meaning that some groups are over- or under-represented. The survey weights in NHANES account for unequal probability of selection as well as for nonresponse. Both the 2003-2004 and 2005-2006 NHANES data provide sample weights for each individual. 
Nusser et al. \cite{nusser} uses the empirical cumulative distribution function to create an equal weight sample from the raw data. Using the notation of \cite{nusser}, the empirical cumulative distribution function for the raw measurement of risk factor $rf$ is calculated as:

\begin{align}
	\hat{F}_{rf}(a) &= \sum_{i=1}^{n} w_i I_{Y^*_{i,rf}}(a), \\
I_{Y^*_{i,rf}}(a) &= 
    \begin{cases}
 1 \text { if } Y^*_{i,rf} \leq a, \\
  0 \text{ otherwise.}
   \end{cases}
\end{align}

where $w_i$ is the survey weight corresponding to individual $i$ and $Y_{i,rf}^*$ is the original measured value for individual $i$ and risk factor $rf$. The adjusted value $Y_{i,rf} = \hat{F}_{rf}^{-1} \left[ (1/n)(s_i-0.5) \right]$, where $s_i$ is the rank of $Y_{i,rf}^*$. This adjustment is made for each of the seven MetS risk factors on all individuals, and this adjustment keeps the support of the original data, which for each risk factor is $(0,\infty)$.


\section{Measurement Error and Regression Models}

\subsection{Measurement Error Model (MEM)}

We denote \emph{usual minutes} of MVPA per day for individual $i$ as $t_i$, in the notation of \cite{nusser}:

\begin{align}
t_i &\equiv E(T_{ij}|i) ,
\end{align}

where $T_{ij}$ is the true number of minutes in MVPA on day $j$ for individual $i$. Neither $t_i$ nor $T_{ij}$ is  directly observed, rather an accelerometer measurement of daily minutes in MVPA, $W_{0ij}$, is measured and  adjustments described in Section \ref{sec:dayadjustments} are applied to get $W_{1ij}$. This observation is contaminated with measurement error through error in the measurement device and day to day variability in physical activity.

Because not everyone participates in MVPA every day, the model needs to allow for both positive and zero minutes of MVPA in a day. In the statistics literature, this is often referred to as \emph{semicontinuous} data. Statistical methods for semicontinuous data were considered as early as 1955 by \cite{aitchison1955} and have been developed through applications such as health sciences and services \cite{neelon2016a,neelon2016b,tooze,smith2014}, nutrition \cite{kipniszeros1,kipniszeros2,zhang1,zhang2,nci,olsen}, manufacturing \cite{lambert1992}, economics \cite{cragg1971}, and climate science \cite{kumar2017}.

\subsubsection{MEM Assumptions}
\label{sec:assumptions}

Our model follows the frameworks of selected works in nutrition \cite{olsen, nci, kipniszeros1} where the MEM contains   two components: the probability of an individual participating in MVPA, and given participation in MVPA, for how many minutes. Individual $i$'s usual minutes in MVPA can be decomposed as:

\begin{align}
	t_i &= E(T_{ij}\times I(T_{ij}>0)|i) = E(T_{ij}|T_{ij}>0,i) P(T_{ij}>0|i) = A_i \pi_i,
\end{align}  

where $A_i$ is the usual minutes of MVPA on a day where individual $i$ participated in MVPA, and $\pi_i$ is the probability that individual $i$ participates in $>0$ minutes of MVPA on any given day. Like \cite{nci} and \cite{kipniszeros1}, we assume: i) MVPA is reported by the accelerometer if and only if the individual actually participated in MVPA,  ii) the accelerometer is unbiased for usual minutes of MVPA on days of MVPA participation. More formally:

\begin{align}
	E(W_{1ij}|i) = E(W_{1ij}|W_{1ij}>0,i) P(W_{1ij}>0|i) \overset{assume}{=} A_i \pi_i.
	\label{ti}
\end{align}

\subsubsection{MEM Preliminaries}
\label{sec:preliminaries}

The MEM assumes the measurement error of MVPA is independent of true minutes of MVPA. Additionally, our MEM will using a Normal distribution for (positive) minutes of MVPA. Taking the fourth root of the observed value makes these assumptions reasonable. Let $W_{ij}=W_{1ij}^{1/4}$. 

MVPA data from accelerometers from NHANES were collected across seven continuous days for each individual, so autocorrelation is an issue. Preliminary analysis  showed an AR(1) model, with parameters $\phi_{g(i)}$,  to sufficiently capture error autocorrelations, where $g(i) = 1$ if individual $i$ is $<65$ years old and 2 otherwise.

The measurement error is assumed to be independent of minutes in MVPA, but it is possible that measurement error varies based on demographic variables. Preliminary analysis also showed a cubic trend with age in MEM residuals (after including AR(1) model of previous paragraph; plot included in supplemental material). 
The cubic model in equation \eqref{varfcn} was fit with OLS to account for differences in measurement errors by age:

\begin{align}
	var(e_{ij}) &= \xi^2_i(\boldsymbol{\delta}) = \delta_0 + \delta_1age_i + \delta_2 age_i^2 + \delta_3 age_i^3 + \nu_{ij}, 	\label{varfcn} \\
	\nu_{ij} &\sim N(0,\sigma^2_{\nu}),
\end{align}

where $e_{ij}$ are the residuals of the preliminary analysis model with an AR(1) error structure. Because we are not interested in making inference on these parameters, the estimate of measurement error variance for individual $i$, $\xi_i^2(\hat{\boldsymbol{\delta}}), \hat{\boldsymbol{\delta}}=(\hat{\delta}_0,\hat{\delta}_1,\hat{\delta}_2,\hat{\delta}_3)$, is used as a plug-in estimate in the full MEM.

\subsubsection{MEM Specification}
\label{sec:specification}

Let ${\bf W}_{iC(i)}$ be a vector of length $m_i$ containing positive values  of $W_{ij}$ for individual $i$. The function $C(i)$ returns the day indices for which $W_{ij}>0$.  For example, if $W_{ij} >0$ for $j=1,2,5$, then $C(i) = (1,2,5)$ and ${\bf W}_{iC(i)}=(W_{i1},W_{i2},W_{i5})'$. Combining assumptions and preliminaries from Sections \ref{sec:assumptions} and \ref{sec:preliminaries}, respectively, the complete MEM is:

\begin{align}
\begin{split}
	I(W_{ij} > 0)|\pi_i &\overset{ind}{\sim} \text{Bernoulli}(\pi_i), \\ 
	\pi_i|b_{1i},{\bf Z}_i & =  logit^{-1}({\bf Z}'_i\boldsymbol{\alpha} + b_{1i}), \\
	{\bf W}_{iC(i)}|b_{2i},{\bf Z}_i &\overset{ind}{\sim} N \left(  {\bf Z}'_i\boldsymbol{\beta} + b_{2i},\Sigma_{}(\phi_{g(i)},\xi_i^2(\hat{\boldsymbol{\delta}})) \right), \\ 
	b_{1i},b_{2i} & \overset{ind}{\sim} N({\bf 0}, \Sigma_b),
	\label{mem}
\end{split}
\end{align}

where   $b_{1i}$, $b_{2i}$ are person level random effects with covariance matrix $\Sigma_b$. 
The covariates are included in the vector ${\bf Z}_{i}$ and its corresponding regression parameters of the two-part model are given by $\boldsymbol{\alpha}$ and $\boldsymbol{\beta}$. The covariance of  ${\bf W}_{iC(i)}$ is denoted by $\Sigma_{}(\phi_{g(i)},\xi^2(\hat{\boldsymbol{\delta}}))$ and is an $m_i \times m_i$ covariance matrix with elements  in the $kl^{th}$ position, $\sigma_{i,kl}$, defined as:

\begin{align}
	\sigma_{i,kl} &= \xi_i^2(\hat{\boldsymbol{\delta}}) \phi_{g(i)}^{|C(i)[k]-C(i)[l]|},
\end{align}

 where $C(i)[k]$ denotes the $k^{th}$ element of $C(i)$ and $\xi_i^2(\hat{\boldsymbol{\delta}})$ is the plug-in estimate defined in equation \eqref{varfcn}. Recall $g(i) = 1$ if individual $i$ is under 65, and 2 otherwise.

\subsubsection{Estimating Usual Minutes in MVPA}

An unbiased estimate of $t_i$, is obtained with a Taylor Series approximation using the relation in equation \eqref{ti},

\begin{align}
	A_i &=  E(W_{ij}^4|W_{ij}>0,i)	\label{condexp} \\
	&\approx (  {\bf Z}'_i  \boldsymbol{\beta}+ b_{2i} )^4 + 6 \sigma_{i,jj} (  {\bf Z}'_i \boldsymbol{\beta}+ b_{2i} )^2. \label{condexpappx}
\end{align}

Combining the assumptions in equation \eqref{ti} with the approximation in equation \eqref{condexpappx},  usual number of minutes in MVPA for individual $i$, $t_i$,  is given by:

\begin{align}
	{t_i} &= {A_i} {\pi_i} \approx  logit^{-1}({\bf Z}'_i\boldsymbol{\alpha} + b_{1i}) \left(  (  {\bf Z}'_i \boldsymbol{\beta} + b_{2i} )^4 + 6 \sigma_{i,jj} (  {\bf Z}'_i \boldsymbol{\beta} + b_{2i} )^2 \right).
	\label{tiestimate}
\end{align}

\subsection{MetS Risk Factor Model (RFM)}

\begin{figure}[h]
\centering
\includegraphics[width=0.6\textwidth]{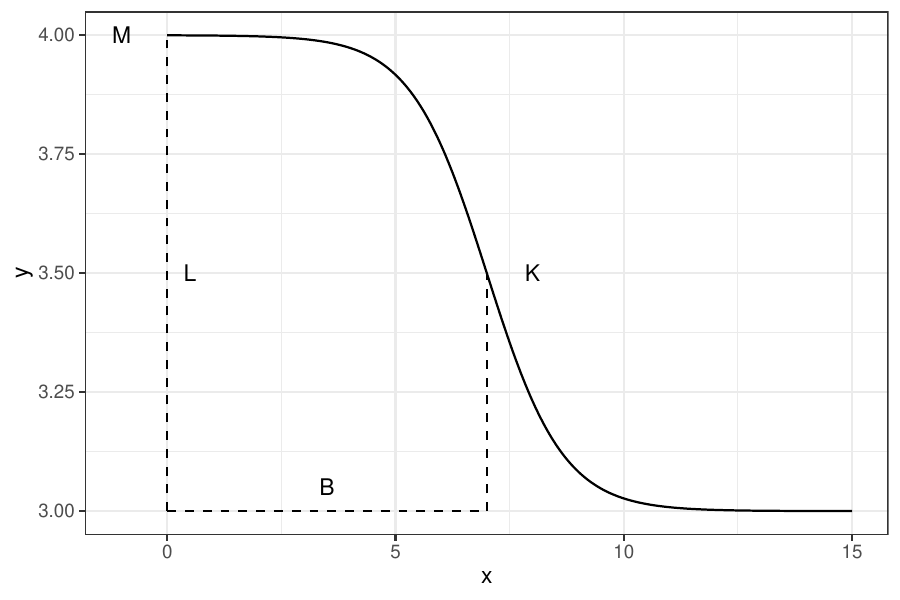}
\caption{Nonlinear function for relationship of MVPA with waist circumference, glucose, triglycerides, and systolic blood pressure.}
\label{nonlinear}
\end{figure}

We denote the vector of MetS risk factors adjusted for sample weights for individual $i$ by ${\bf Y}_{i}$, which has elements $Y_{i,wst}$  (waist circumference in centimeters), $Y_{i, glu}$ (log glucose (mg/dL)), $Y_{i,tri}$ (log triglycerides (mg/dL)), $Y_{i, ldl}$ (LDL cholesterol (mg/dL)), $Y_{i, hdl}$ (HDL cholesterol (mg/dL)), $Y_{i, bps}$ (systolic blood pressure (mm Hg)), and $Y_{i, bpd}$ (diastolic blood pressure (mm Hg)). The relationship between $t_i$ (obtained using equation \eqref{tiestimate}) and ${\bf Y}_{i}$ is modeled with the risk factor regression model (denoted RFM):

\begin{align}
	E({\bf Y}_{i}|t_i) &= {\bf m}(\boldsymbol{\gamma};t_i).
	\label{regmodel}
\end{align}

where ${\bf m}(\cdot)$ contains the functional form for each of the seven risk factors. Based on Figure \ref{scatter} and the loess fit, waist circumference, log glucose, log triglyceride, and systolic blood pressure all appear to have a similar, upside down S shape. The function:

\begin{align}
	f(x) &= M - \frac{L}{1+e^{-K(x-B)}},
	\label{nonlinearfcn}
\end{align}

has this form. Let $m_{wst}(),m_{glu}(),m_{tri}()$, and $m_{bps}()$ denote the functions with the form in equation \eqref{nonlinearfcn}. Figure \ref{nonlinear} gives an example of equation \eqref{nonlinearfcn} with parameter values: $M=4, L=1, B=7, K=1.2$. 
One of the benefits of this functional form is that the parameters are interpretable: $M$ is the limit of $f(x)$ as $x$ goes to $-\infty$,  $M-L$ is the limit of $f(x)$ as $x$ goes to $\infty$, $B$ is the inflection point of the curve with respect to the x-axis variable, and $K$ is the rate of change. LDL, diastolic blood pressure, and HDL appear to have a fairly linear relationship so $m_{ldl}()$, $m_{bpd}()$, and  $m_{hdl}()$ have linear form. The regression parameters in equation \eqref{regmodel} $\boldsymbol{\gamma}$ are either (L,K,B), or a slope, depending on the MetS risk factor.

Because all the MetS risk factors are characteristics within an individual, there will be dependencies between the errors for each risk factor regression within an individual. Moreover, the common assumption of Normality of regression errors is not reasonable since MetS factors have skewed distributions.  Zellner \cite{zellner} proposed SUR as a generalization of a multivariate linear regression where each dependent variable could have potentially different explanatory variables and different relationships between those variables. Furthermore, errors are assumed to be correlated across the dependent variables. Galimberti et al. \cite{galimberti} proposed using a mixture of Normal distributions to model the errors in SUR in the linear case estimated with maximum likelihood. We extend this model to the  nonlinear case and adopt a Bayesian framework. The proposed model uses a mixture of $H$ Normal distributions for the seven dimensional error distribution. The $n$ seven-dimensional error vectors $\boldsymbol{\epsilon}_i$ are assumed to be independent and identically distributed. The RFM can be written as:

\begin{align}
\begin{split}
	{\bf Y}_{i}|(\zeta_i = h) &= \boldsymbol{\lambda}_h +  {\bf m(}t_i, \boldsymbol{\gamma}) + \boldsymbol{\epsilon}_{i}, \\
	\zeta_i | {\bf p} &\overset{ind}{\sim} \text{Cat}(H, {\bf p}), \\
	\boldsymbol{\epsilon}_{i} &\overset{ind}{\sim}  N({\bf 0}, \Sigma_{m,h}), 
	\label{nonlinearsur}
\end{split}
\end{align}

where $\zeta_i$ is a latent class variable indicating which component of the mixture observation $i$ is located,  Cat($H,{\bf p}$) refers to the categorical distribution with probabilities ${\bf p} = ({p_1,p_2,...,p_H}$) corresponding to the H components of the mixture, and $\boldsymbol{\lambda}_h$ is the component intercept for each risk factor. 
 The 7$\times$7 matrix $\Sigma_{m,h}$ is the covariance matrix for the $h$th component of the mixture (the subscript $m$ is only to distinguish it from equation \eqref{mem}). 
In a slight change of notation, $\boldsymbol{\gamma}_0$ represents the overall intercept vector for the vector of functions ${\bf m}(\cdot)$.  As in \cite{galimberti}, the overall intercept vector is given by:
\begin{align}
	\boldsymbol{\gamma}_0 &= \boldsymbol{\lambda}_1 - \sum_{h=1}^{H} p_h (\boldsymbol{\lambda}_1-\boldsymbol{\lambda}_h).
\end{align}

\subsection{Complete Likelihood}

Let $p()$ denote a probability density/mass function. The likelihood for individual $i$ can be written as:

\begin{align}
	L(\boldsymbol{\theta}|{\bf W}_i,{\bf Y}_i,{\bf Z}_i) &= \int \int p({\bf Y}_i|t_i,\boldsymbol{\theta}_{RFM}) p(\zeta_i|{\bf p}) \left\{ \prod_{j=1}^{J} p(I({ W_{ij} >0}) | {\bf Z}_i, b_{1i},\boldsymbol{\theta}_{{MEM}_1}) \right\} \\
 &\times p({\bf W}_{iC(i)}|{\bf Z}_i, b_{2i}, \boldsymbol{\theta}_{{MEM}_1}) p(b_{1i},b_{2i}|\boldsymbol{\theta}_{{MEM}_2}) db_{1i} db_{2i},
\end{align}

where $ p({\bf Y}_i|t_i,\boldsymbol{\theta}_{RFM})$ is given by equation \eqref{nonlinearsur}, $p(I({ W_{ij} >0}) | {\bf Z}_i, b_{1i},\boldsymbol{\theta}_{{MEM}_1})$, $p({\bf W}_{iC(i)}|Z_i, b_{2i}, \boldsymbol{\theta}_{{MEM}_1})$, and $p(b_{1i},b_{2i}|\boldsymbol{\theta}_{{MEM}_2})$ are given by equation \eqref{mem}. We let $\boldsymbol{\theta}_{RFM} = (\boldsymbol{\gamma},\{ \boldsymbol{\lambda} \}_{h=1}^H, \{ \Sigma_{m,h} \}_{h=1}^H, \zeta_i,{\bf p})$, $\boldsymbol{\theta}_{{MEM}_1} =(\boldsymbol{\alpha},\boldsymbol{\beta},\boldsymbol{\phi})$, and  $\boldsymbol{\theta}_{{MEM}_2} = (\Sigma_b)$, $\boldsymbol{\theta}_{{MEM}} = (\boldsymbol{\theta}_{{MEM}_1} ,\boldsymbol{\theta}_{{MEM}_2}) $  and $\boldsymbol{\theta}=(\boldsymbol{\theta}_{{RFM}},\boldsymbol{\theta}_{{MEM}})$. Assuming independence across individuals, the full likelihood is:

\begin{align}
	L(\boldsymbol{\theta}|{\bf W,Y,Z}) &= \prod_{i=1}^{n} L(\boldsymbol{\theta}|{\bf W}_i,{\bf Y}_i,{\bf Z}_i).
\end{align}

Priors for parameters are assumed to be independent \emph{a priori}, and where possible, conjugate. The priors for the MEM were chosen to be non-informative.  Because the MetS risk factor regression model has interpretable parameters, we used data from the 2001 NHANES to construct priors. The priors and details on how the informative priors were constructed are included in Appendix A  for brevity. Other years' data could be used in lieu of or in conjunction with the 2001 NHANES to construct priors, but we suspect changes in the results would be relatively minor since NHANES is a large population-level sample. If a prior was constructed from a non-representative sample in regards to this study, for example marathon runners, the prior would likely skew the results away from their intended population.

\subsection{Estimation}

It is computationally expensive to run the MCMC on the joint MEM-RFM at once, so estimation is done for the two components of the model in sequence. The first stage runs the MCMC for the MEM in equation \eqref{mem}. The posterior distribution for ${\bf t}=(t_1,t_2,...,t_n)'$, $p({\bf t| W,Z})$, is obtained using $p( \boldsymbol{\theta}_{MEM}|{\bf W,Z})$ with equation \eqref{tiestimate}. The second stage runs the MCMC for the regression model in equation \eqref{nonlinearsur}, using $p({\bf t| W,Z})$ as the true MVPA. Each MCMC iteration in the second stage draws a ${\bf t}$ from $p({\bf t| W,Z})$ randomly, and treats this vector as known. Algorithm \ref{alg:est} details this two-stage procedure.

Note this approach does not require having both MVPA and MetS risk factor data for all individuals. Individuals who only have MVPA data can help estimate the parameters of the MEM. The RFM is then estimated with a subset of these individuals who also have full data on risk factors.

\begin{algorithm}
\caption{Estimation Procedure of Full Model}\label{alg:est}
\begin{algorithmic}
\Require $N_{MEM}$: Number of MCMC draws for MEM
\Require $N_{RFM}$: Number of MCMC draws for RFM
   \State ${\bf W} \gets $ Take fourth root of ${\bf W}_{1}$
   \State $p(\boldsymbol{\theta}_{MEM}^{}|{\bf W}^{},{\bf Z}) \gets $ Draw $N_{MEM}$ samples from ME posterior via \hspace*{4.5cm} MCMC  in Stan
	\State $p({\bf t}^{}|{\bf W}^{},{\bf Z})  \gets $ Compute posterior of \emph{usual} daily minutes of MVPA using \hspace*{4.5cm} $p(\boldsymbol{\theta}_{MEM}^{}|{\bf W}^{},{\bf Z})$ and equation \eqref{tiestimate}
\For{$\ell \text{ in } 1:N_{RFM}$}
\State ${\bf t}_{\ell}^{} \gets $ Sample one vector from $p({\bf t}^{}|{\bf W}^{},{\bf Z})$
\State $p(\boldsymbol{\theta}_{RFM}^{}|{\bf t}_{\ell}^{},{\bf Y,Z}) \gets $ Draw 1 sample from MCMC using sampled $ {\bf t}_{\ell}^{}$
\EndFor
\State $p(\boldsymbol{\theta}_{RFM}^{}|{\bf t},{\bf Y,Z}) \gets \{p(\boldsymbol{\theta}_{RFM}^{}|{\bf t}_{\ell}^{},{\bf Y,Z})\}_{\ell=1}^{N_{RFM}}$
\end{algorithmic}
\end{algorithm}

This approach is similar to regression calibration \cite{nonlinear}, but the major difference is for every MCMC draw from the posterior distribution, there is a new estimate of $t_i$ for every individual. Regression coefficients in equation \eqref{regmodel} are estimated using these $t_i$ values. This procedure automatically accounts for the variability in the MEM regression coefficients in equation \eqref{tiestimate} used to calculate $t_i$. Regression calibration uses one set of $\hat{t}_i$ values  computed from equation \eqref{tiestimate}, and then those $\hat{t}_i$ are plugged into equation \eqref{regmodel}. The latter approach requires using the bootstrap or other methods to adjust standard errors, whereas our Bayesian approach does it all together.

\section{Results}

Posterior distributions of the  parameters of the MEM in equation \eqref{mem} are estimated  using Stan \cite{carpenter}. To approximate the posterior distribution of the MEM, we generated eight chains each of length 2000, and used the first 1000 as burn-in in Stan. Trace plots and Gelman-Rubin diagnostics showed good mixing. 

To estimate the parameters of the RFM in equation \eqref{nonlinearsur}, we wrote an MCMC in \texttt{R} and C++.  Using the estimated $p({\bf t|W,Z})$  from Stan, we used a Gibbs sampler to generate samples from the joint posterior distribution of all RFM parameters. The intercepts for all components of the mixture distribution for the residuals, covariance matrices of the mixture distribution, and the mixing weights, all had conjugate full conditional distributions. The remaining regression parameters are drawn using a Metropolis step with Normal random walk proposals. Full conditionals for these parameters are given in the supplemental material.  We tune the proposal covariance matrix using the recommended $2.4^2 Var(\theta|y)/d$ in \cite{bda}, pg. 290, where $d$ is the dimension of $\theta$. Because we are using a mixture model for the MetS risk factors, there can be a symmetry in the model parameters, leading to a problem known as ``label switching" \cite{stephens}. In brief, the MCMC-algorithm does not keep track of which parameters correspond to which mixture component across MC iterations (e.g. mixture component $1$ at MC iteration $\ell$ might be mixture component $2$ at MC iteration $\ell+1$), meaning mixture component $h$ doesn't have a consistent meaning across MC samples, making it difficult to distinguish after sampling. We use the approach proposed by Stephens \cite{stephens} who uses a decision theoretic approach to address the problem of label switching, where the labels are found which minimize the Kullback-Leibler divergence between average component weights for the entire MCMC run and  the classification for each MCMC iteration. This method is implemented in the R package \texttt{label.switching} \cite{papastamoulis}.  
For the regression model, we generated three chains of length 1,000,000, using the first 10,000 as burn-in, and thinning every 5 iterations to save on memory and reduce auto-correlation. The chain showed no signs of nonconvergence. 
The value of $H$ is considered to be fixed in equation \eqref{nonlinearsur}, so we pick the number of mixing components that minimizes the model Deviance Information Criterion \cite{spiegelhalter}, which is five for this problem.

\subsection{RFM Parameter Estimates}

Regression parameter estimates for the nonlinear and linear functions are given in Tables \ref{nonlinearestimp} and \ref{linearestimp}, respectively, together with 95\% credible sets created from the combined posterior. 
Recall that the parameters of interest in the nonlinear function are $L$ and $B$, where $L$ is the benefit that can be gained from participating in MVPA and $B$ is the inflection point, which can be thought of as the minimum number of minutes of MVPA required to maximize benefits toward the specific MetS risk factor; there are increasing returns for MVPA up until that point.

With elevated time in MVPA, the individuals in the sample showed  an aggregate  lowering of waist circumference by 11 cm, or $B$, (recall definitions of nonlinear function from equation \eqref{nonlinearfcn} and Figure \ref{nonlinear}),  systolic blood pressure by 19 mm Hg, glucose levels by 14 mg/dL ($e^{\gamma_0} - e^{\gamma_0-L}$), and triglyceride levels  by 26 mg/dL, on average, when considering individuals who participate in 0 usual minutes of MVPA compared to those who participate in MVPA regularly (defined as 2$\times B^4$ for each respective risk factor modeled nonlinearly).  All of these effects are statistically significant and practically important. The amount of time that is relevant for each risk factor depends on $B$. The nonlinear function's inflection point, $B$, range in value on the transformed scale from 1.3 for Glucose to 2.5 for log triglyceride, for fourth root of minutes, which translates to about 3 minutes to 40 minutes of MVPA per day. This is a large range to achieve the optimal tradeoff between health benefits and time. The data suggest further health benefits beyond  this reflection point, but with diminishing returns. The low amount of time required for improvement in Glucose suggests there may be additional factors at play since 3 minutes is a low amount of MVPA to have a practical difference. The authors note, these are observational results only, but provide insight into population level trends. Triglyceride levels and waist circumference on the other hand, seem to require much higher levels, at the population level, all else equal. 

The estimated slopes in the linear regressions are not entirely in line with our expectations. The coefficients for HDL and diastolic blood pressure are negative and positive, respectively. We would have expected HDL to increase with increased MVPA and vice versa for diastoic blood pressure, although this doesn't necessarily appear to be the case in Figure \ref{scatter} either. It is possible these effect is confounded with other, non-measured variables. 
\begin{table}[ht]
\centering
\caption{RFM parameter estimates in the nonlinear functions.}
\scalebox{0.8}{

\begin{tabular}{ll|ll}
  \hline
Parameter & MetS RF & Post Mean & 95\% CI \\ 
  \hline
$\gamma_0$ & Waist (cm) & 101.09 & (98.81,103.67) \\ 
   & log Glucose (log mg/dL) & 4.71 & (4.65,4.78) \\ 
   & log Triglyceride (log mg/dL) & 4.77 & (4.73,4.83) \\ 
   & Sys Blood Press (mm Hg) & 137.17 & (134.46,139.44) \\ 
   \hline
L & Waist & 11.36 & (9.77,12.43) \\ 
   & log Glucose & 0.14 & (0.09,0.21) \\ 
   & log Triglyceride & 0.25 & (0.08,0.75) \\ 
   & Sys Blood Press & 19.22 & (18.39,19.99) \\ 
   \hline
K & Waist & 1.64 & (0.99,2.58) \\ 
   & log Glucose & 2.56 & (1.86,3.28) \\ 
   & log Triglyceride & 3.68 & (3.00,4.53) \\ 
   & Sys Blood Press & 3.02 & (2.16,4.16) \\ 
   \hline
B & Waist & 2.31 & (1.69,2.92) \\ 
   & log Glucose & 1.27 & (0.83,1.70) \\ 
   & log Triglyceride & 2.54 & (1.90,3.37) \\ 
   & Sys Blood Press & 1.44 & (1.17,1.73) \\ 
   \hline
\end{tabular}}

\label{nonlinearestimp}
\end{table}

\begin{table}[ht]
\centering
\caption{RFM parameter estimates in the linear functions.}
\scalebox{0.8}{

\begin{tabular}{ll|ll}
  \hline
Parameter & MetS RF & Post Mean & 95\% CI \\ 
  \hline
$\gamma_0$ & LDL (mg/dL) & 128.81 & (126.46,131.14) \\ 
   & Dias Blood Press (mm Hg) & 64.81 & (62.92,66.68) \\ 
   & HDL (mg/dL) & 62.87 & (61.63,64.10) \\ 
   \hline
$\gamma_1$ & LDL & -5.91 & (-6.41,-5.39) \\ 
   & Dias Blood Press & 2.63 & (1.75,3.53) \\ 
   & HDL & -3.32 & (-3.70,-2.91) \\ 
   \hline
\end{tabular}}

\label{linearestimp}
\end{table}

We also considered the effect of the MEM on the RFM by estimating the RFM using the raw measurements $W_{ij}$. The analysis uses the same 1039 individuals on the full MEM-RFM model. All other components of the model and estimation procedures are the same. Overall, results follow the same trends. Some notable differences include the estimate of $B$ for triglycerides, all estimates of $K$, and the slope for diastolic blood pressure. The effect of this difference in the estimate of $B$, is 40 minutes to 10 minutes for the model not accounting for measurement error. The effect of lower $K$  is a more spread out effect of MVPA on each risk factor. Full results for the RFM using the raw measurements $W_{ij}$ are in the supplemental material.

\subsection{Probability of MetS}

\begin{figure}
\centering
\includegraphics[width=5in,height=2.5in]{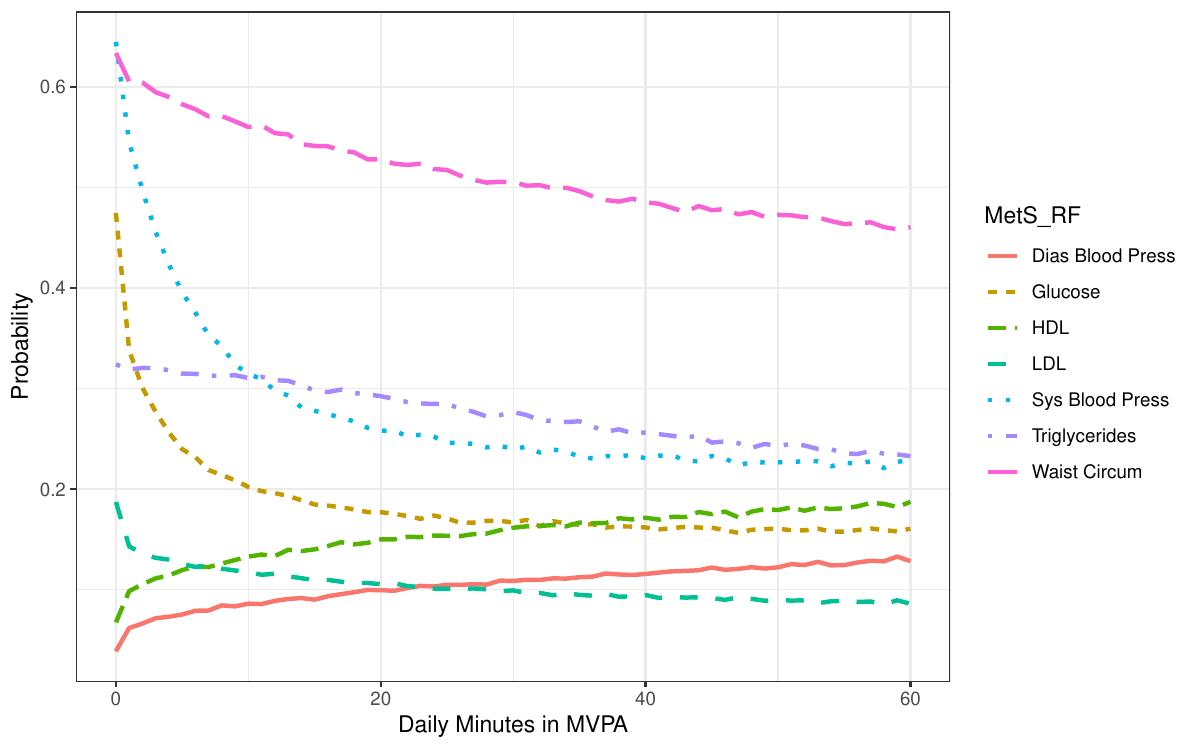}
\caption{Probability an individual will have a high level for each MetS risk factor as function of daily minutes in MVPA based on observational, cross-sectional data.  }
\label{indmetsrfimp}
\end{figure}

We can still explore population level relationships between MetS risk factors and MVPA by calculating the proportion of the population with high levels of MetS risk factors as a function of minutes in MVPA. This can be computed directly from the posterior predictive distribution of the MetS risk factors conditioned on minutes in MVPA. The posterior predictive distribution is:

\begin{align}
 p(\tilde{{\bf Y}}|{\bf W,Y,Z}) &= \int \int p(\tilde{{\bf Y}}|\tilde{{\bf t}},\boldsymbol{\theta}_{RF}) p(\boldsymbol{\theta}|{\bf W}, {\bf Y}, {\bf Z}) d b_{1} d  b_{2} d\boldsymbol{\theta},
	\label{ypostpred}
\end{align}

where variables with a tilde (eg. $\tilde{Y}$), denote variables for a new observation and $p(\boldsymbol{\theta}|{\bf W}, {\bf Y}, {\bf Z})$ is the posterior distribution of all the model parameters. Figure \ref{indmetsrfimp} shows the probability of an individual having a high level of each risk factor as a function of minutes in MVPA ($P(\tilde{Y}_{rf} > y_{rf}|\tilde{t})$, where $y_{rf}$ is the ``high level" of risk factor $rf$ explained in Section 1). There is a fast drop in probability for both systolic blood pressure and glucose, as suggested by the values of parameter estimates.  Notice that even with 60 minutes of MVPA a day on average, 46\% of the represented population is still expected to have a  large waistline. This suggests that factors other than exercise could have a major impact on one's weight and waistline. 

\begin{figure}
\centering
\includegraphics[width=5in,height=2.5in]{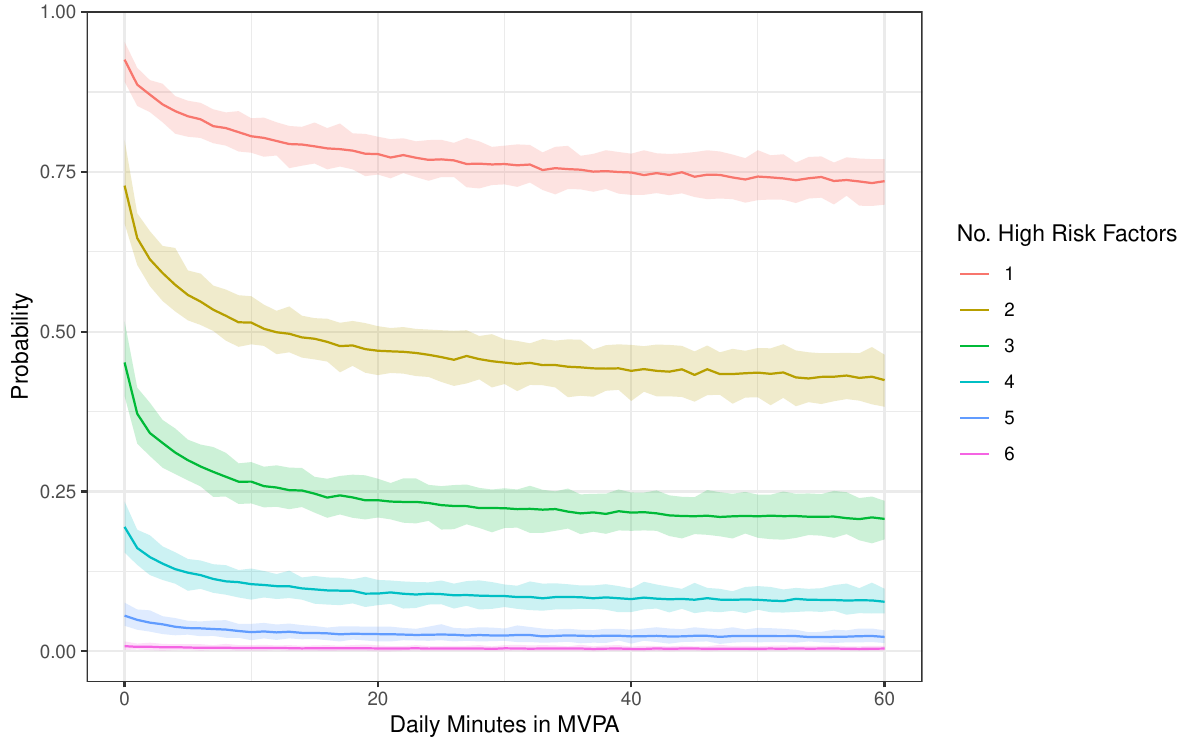}
\caption{Probability of exhibiting $R=1,2,...6$ or more elevated levels of MetS risk factor as a function of daily minutes in MVPA.  }
\label{metsprob}
\end{figure}

MetS is diagnosed when a person exhibits multiple elevated risk factors.  Figure \ref{metsprob} shows the probabilities that an individual exhibits $R$ or more risk factors elevated levels, for $R$=1,2,...,6, along with pointwise 95\% credible intervals, constructed using the  posterior predictive distribution defined in equation \eqref{ypostpred}.
It is  interesting to note that even for those who participate in 60 minutes of MVPA daily on average, there is still a nearly 70\% chance of having at least one elevated MetS risk factor. This suggests the diminishing returns of MVPA in relation to MetS risk factors.

\subsection{Residual Analysis}

To assess the fit of model equation \eqref{nonlinearsur}, we calculate standardized residuals and perform residual analysis. We let the residual vector for individual $i$ be:

\begin{align}
	{\bf r}_i &= \frac{{\bf Y}_{i} - E({\bf Y}_i)}{\sqrt{var({\bf Y}_i)}},
	\label{stdresid}
\end{align}

None of the residual plots (included in the supplemental material) for the MetS risk factors appear to show  nonconstant variance or other trends, when plotted against the predicted usual minutes of MVPA, indicating a good model fit.

\section{Discussion}

Metabolic Syndrome is a condition affecting millions of people and has multiple causes, for which low levels of  physical activity is thought to be one. To understand the relationship between physical activity and MetS, one must first be able to measure physical activity. Measures of MVPA can contain device and user errors and biases, as well as day to day MVPA variability. Although MetS diagnosis is binary, MetS itself is composed of several risk factors which are physiological measurements, meaning modeling the relationship between MVPA and MetS may be more informative on the granular, measurement level of MetS. Furthermore, each of the risk factors may have a different relationship with MVPA, and there is dependence across these risk factors within an individual. Accurately accounting for these measurement errors, varying relationships, and correlations is important to fully understand the effects of MVPA on MetS.

This article presents a two-stage model to help answer how MVPA affects MetS. This includes a measurement error model for time in MVPA which accounts for errors and biases in raw measurements. The MEM is flexible and includes a covariance structure that has time dependence built in to account for consecutive day monitoring and heterogeneity in the variances as functions of model covariates. Once these estimates of usual minutes in MVPA are obtained, they can then be passed to a risk factor regression model to explore their relationship with each risk factor. The RFM model lets each risk factor have different relationships with MVPA, and it accounts for correlation of residuals for different risk factors within an individual. The residuals themselves are modeled as a mixture of Normals, creating a flexible error distribution which eliminates stringent modeling assumptions. The Bayesian framework makes it easy to connect these two models and correctly propagate uncertainties through and maximize the information gained from the data.

The model is fit to data from the 2003-2006 NHANES. Adjustments for survey weights and weekend effects were made. Data from the 2001 NHANES was used to construct priors for the regression parameters  in the SUR model. Results were mostly consistent with expectations and suggest that time in MVPA can help reduce the risk of developing MetS. At the individual risk factor level, the model suggested waist circumference requires higher levels of MPVA to see benefits compared to others. All risk factors saw practical and statistical relationships in the sample, on average, at higher levels of MVPA. Additional benefits are generally seen with more time spent in MVPA. It is worth noting that doing a linear regression on diastolic blood pressure (HDL) on mean, fourth root  time in MVPA by individual, results in positive (negative) coefficients as well, even though it goes against current knowledge. However, this is a cross-sectional observational study meaning these results are only suggestive of causal effects. Even with that caveat, these results can help  prescribe time in  MVPA for future intervention studies to understand the causal relationships with each MetS risk factor. 

There are some limitations to this study. As already mentioned, causal conclusions are not possible. There was not consideration for individuals taking medicine for either blood pressure or cholesterol. NHANES does ask if participants are taking a prescription for hypertension, but 73\% are missing values, and NHANES only asks if individuals are told to take a prescription for cholesterol and 78\% are missing values. Third, participants in the sample could be on a physical activity intervention, where they started increasing physical activity levels, but the long run benefits have not been achieved yet. We also note that there have been different findings based on whether one is participating in aerobic or resistance based exercise, of which both could be at the moderate to vigorous intensity, and our data does not distinguish between the two. There was also a signifiant amout of missing accelerometry data. This limited the number of individuals we could include in MEM and as a result, the RFM. Many of these limitations have to do with the data collection aspect, which is often messy.

\section*{Acknowledgements}

The authors gratefully acknowledge Dr. Kevin W. Dodd from NCI with whom we had fruitful discussions and who generously shared code for some of the initial processing of the NHANES data.

\section*{Disclosure Statement}
No potential conflict of interest was reported by the authors. 

\section*{Funding}
This work was supported by National Institutes of Health Grant number HL091024. Sandia National Laboratories is a multimission laboratory managed and operated by National Technology and Engineering Solutions of Sandia, LLC, a wholly owned subsidiary of Honeywell International Inc., for the U.S. Department of Energy's National Nuclear Security Administration under contract DE-NA0003525. This paper describes objective technical results and analysis. Any subjective views or opinions that might be expressed in the paper do not necessarily represent the views of the U.S. Department of Energy or the United States Government.

\section*{Data Availability Statement}

The data that support the findings of this study are openly available in https://wwwn.cdc.gov/nchs/nhanes/default.aspx.

\bibliographystyle{tfs}

\bibliography{references}

\section{Appendices}
\appendix

\section{Priors}

The priors for the MEM parameters are given below:

\begin{multicols}{2}
\noindent 
\begin{align}
\boldsymbol{\beta} &\sim N({\bf 0}_{8},1000 \times I_{8}) \\
\boldsymbol{\alpha} &\sim N({\bf 0}_{8},1000 \times I_{8}) \\
\phi_g &\sim \text{Unif}(-1,1), g=1,2 
\end{align}
\columnbreak
\begin{align}
\Sigma_b & = \boldsymbol{\sigma}_b' \Omega \boldsymbol{\sigma}_b \\
\Omega &\sim  \text{LKJ}(1.0) \\
\sigma_b &\sim \text{Cauchy}^+ (0,1)
\end{align}
\end{multicols}

where LKJ(1.0) is a distribution for correlation matricies proposed by \cite{lewandowski}.  The RFM model has seven components, each with parameter vector  $\boldsymbol{\gamma}$ = ($\boldsymbol{\gamma}_{wst}$, $\boldsymbol{\gamma}_{glu}$, $\boldsymbol{\gamma}_{tri}$, $\boldsymbol{\gamma}_{bps}$, ${\gamma_{bpd,1}}$, ${\gamma_{ldl,1}}$, ${\gamma_{hdl,1}})$. The parameter vectors for the nonlinear regressions $\boldsymbol{\gamma}_{wst},\boldsymbol{\gamma}_{glu},\boldsymbol{\gamma}_{tri},\boldsymbol{\gamma}_{bps}$ each have elements $L,K,B$. For the linear regression components, parameters ${\gamma_{bpd,1}},{\gamma_{ldl,1}},{\gamma_{hdl,1}}$ are the slopes associated with minutes in MVPA and each respective MetS risk factors. The intercept vector for the regression is $\boldsymbol{\lambda}=(\boldsymbol{\lambda}_{wst},\boldsymbol{\lambda}_{glu},\boldsymbol{\lambda}_{tri},\boldsymbol{\lambda}_{bps},\boldsymbol{\lambda}_{bpd},\boldsymbol{\lambda}_{ldl},\boldsymbol{\lambda}_{hdl})$, where each $\boldsymbol{\lambda}_{u}$ is a H dimensional vector.

 We used the 2001 NHANES self-report for physical activity where individuals are categorized their physical activity levels between 1 and 4, 4 being most active. To set prior means for $\boldsymbol{\lambda}$, we took the mean values of the respective MetS risk factors for those individuals who reported the lowest usual physical activity on the self report. Prior means for L were chosen by taking the difference in mean MetS levels for individuals who responded a 4 for physical activity level and those who responded 1. Prior means for B where chosen to  be equal to $20^{1/4}=2.11$ since the Physical Activity Guidelines recommend approximately 20 minutes of activity at a 3.0 MET level per day, on average. Prior means for K were selected to be equal to 3, so that there is little additional benefit beyond 80 minutes per day of MVPA with an inflection point at 2. Prior variances were selected to allow a small amount of mass at 0 for the priors for K and B. Priors for the regression parameters for LDL, HDL, and diastolic blood pressure are selected to be flat. We selected the prior for the covariance matrix of errors and for the component weights to be conjugate. 

The priors for the regression model parameters are given below:
\begin{align}
\boldsymbol{\lambda}_{wst} &\sim N({\bf 98}_H,17^2 I_H) \\
\gamma_{wst,L} &\sim N(7,8^2) \\
\gamma_{wst,K} &\sim N(3,1.5^2) \\
\gamma_{wst,B} &\sim N(2.11,.4^2) \\
\boldsymbol{\lambda}_{bps} &\sim N({\bf 130}_H,7^2 I_H) \\
\gamma_{bps,L} &\sim N(18,5^2) \\
\gamma_{bps,K} &\sim N(3,1^2) \\
\gamma_{bps,B} &\sim N(1.3,1^2) \\
\boldsymbol{\lambda}_{glu} &\sim N({\bf 4.7}_H,.1^2 I_H) \\
\gamma_{glu,L} &\sim N(.16,.08^2) \\
\gamma_{glu,K} &\sim N(3.6,.7^2) \\
\gamma_{glu,B} &\sim N(1.4,1^2) \\
\boldsymbol{\lambda}_{bpd} &\sim N({\bf 0}_H,100^2 I_H) \\
\gamma_{bpd,1} &\sim N(0,100^2) \\
\boldsymbol{\lambda}_{ldl} &\sim N({\bf 0}_H,100^2 I_H) \\
\gamma_{ldl,1} &\sim N(0,100^2) \\
\boldsymbol{\lambda}_{tri} &\sim N({\bf 4.73}_H,.6^2 I_H) \\
\gamma_{tri,L} &\sim N(.12,.4^2) \\
\gamma_{tri,K} &\sim N(4.88,2^2) \\
\gamma_{tri,B} &\sim N(2.11,.4^2) \\
\boldsymbol{\lambda}_{hdl} &\sim N({\bf 0}_H,100^2 I_H) \\
\gamma_{hdl,1} &\sim N(0,100^2) \\
\Sigma_{m,h} &\sim \text{Inv-Wish} (8, I_{7\times 7}), h=1,...H \\
{\bf p} &\sim \text{Dirichlet}({\bf 1}_{H}).
\end{align}

\clearpage

\section*{Supplemental Material}

\subsection*{MEM Residuals vs Age}

\begin{figure}[h]
\centering
\includegraphics[width=8cm,height=5cm]{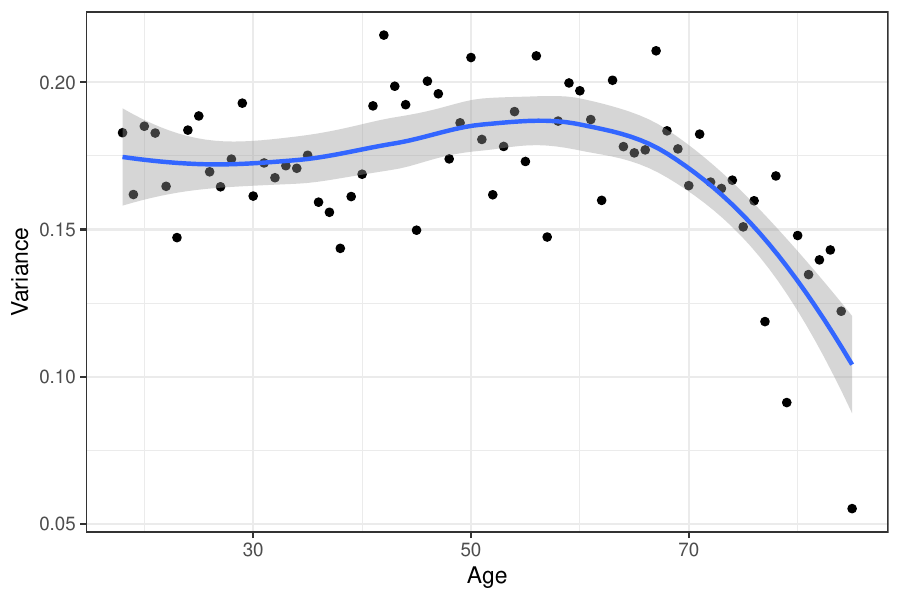}
\caption{Variance by age for residuals of linear mixed effects model for $W_{ij}$ with AR(1) correlation structure. Model was fit using only positive values of $W_{ij}$.}
\label{agevar}
\end{figure}

\clearpage
\subsection*{Parameter Estimates for RFM using observed MVPA (no MEM)}

\begin{table}[ht]
\caption{Parameter estimates in the nonlinear functions for RFM using observed MVPA (no MEM).}
\centering
\begin{tabular}{ll|ll}
\hline
 Parameter & MetS RF & Post Mean& 95\% Credible Interval \\ 
  \hline
$\gamma_0$ & Waist (cm) & 98.44 & (97.15,99.93) \\ 
   & log Glucose (log mg/dL) & 4.68 & (4.65,4.74) \\ 
   & log  Triglyceride (log mg/dL) & 4.81  & (4.76,4.88) \\ 
   & Sys Blood Press (mm Hg) & 135.59  & (133.41,137.65) \\ 
   \hline
L & Waist & 10.27 & (9.81,10.75) \\ 
   & log Glucose &0.11 & (0.07,0.16) \\ 
   & log Triglyceride & 0.17 & (0.09,0.26) \\ 
   & Sys Blood Press & 17.77 & (17.21,18.25) \\ 
   \hline
K & Waist & 3.51  & (3.09,3.91) \\ 
   & log Glucose & 2.89 & (2.44,3.18) \\ 
   & log Triglyceride & 4.50 & (4.34,4.67) \\ 
   & Sys Blood Press & 4.40 & (4.11,4.68) \\ 
   \hline
B & Waist & 2.60 & (2.26,2.92) \\ 
   & log Glucose & 1.05 & (0.55,3.18) \\ 
   & log Triglyceride & 1.78 & (1.25,2.28) \\ 
   & Sys Blood Press & 1.32 & (1.06,1.58) \\ 
   \hline
\end{tabular}

\label{nonlinearest}
\end{table}

\begin{table}[ht]
\caption{Parameter estimates in the linear functions for RFM using observed MVPA (no MEM).}
\centering
\begin{tabular}{ll|ll}
  \hline
Parameter & MetS RF & Post Mean Estimate & 95\% Credible Interval  \\ 
  \hline
$\gamma_0$ & LDL (mg/dL) & 129.25 & (126.97) \\ 
   & Dias Blood Press (mm Hg) & 68.14 & (66.88,69.43) \\ 
   & HDL (mg/dL) & 60.20 & (58.86,61.53) \\ 
   \hline
$\gamma_1$ & LDL (mg/dL) & -6.92 & (-7.37,-6.48) \\ 
   & Dias Blood Press (mm Hg) & 1.03 & (0.40,1.63) \\ 
   & HDL (mg/dL) & -2.14 & (-2.67,-1.61) \\ 
   \hline
\end{tabular}

\label{linearest}
\end{table}

\begin{figure}[h]
\centering
\includegraphics[width=0.8\textwidth]{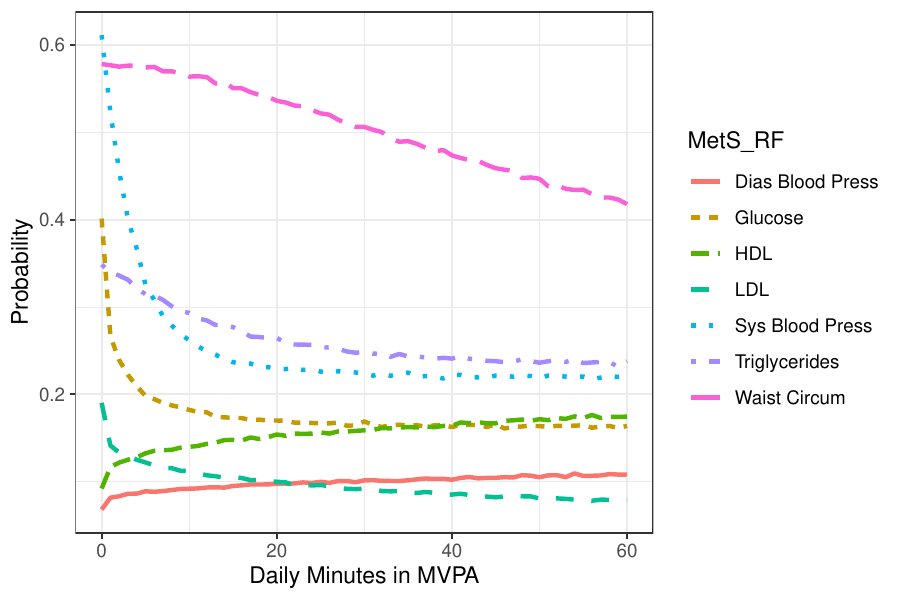}
\caption{Probability an individual will have a high level for each MetS risk factor as function of daily minutes in MVPA for RFM with observed MVPA (no MEM).   }
\label{fig:indmetsrf_noMEM}
\end{figure}

\begin{figure}[h]
\centering
\includegraphics[width=0.8\textwidth]{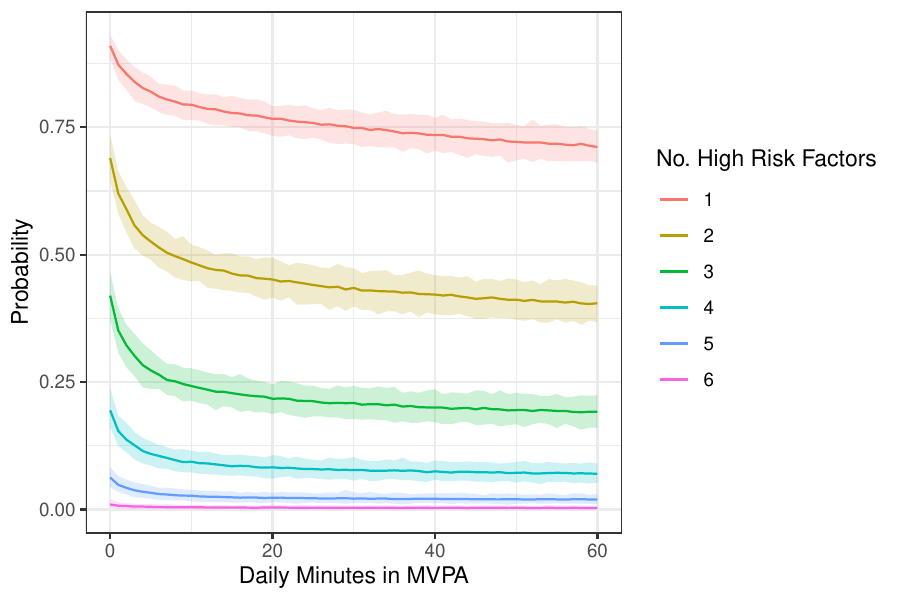}
\caption{Probability of exhibiting $R=1,2,...6$ or more elevated levels of MetS risk factor as a function of daily minutes in MVPA for RFM with observed MVPA (no MEM). }
\label{fig:metsprob_noMEM}
\end{figure}

\clearpage
\subsection*{Standardized Residual Plots for RFM}

\begin{figure}[h]
\centering
\includegraphics[width=0.8\textwidth]{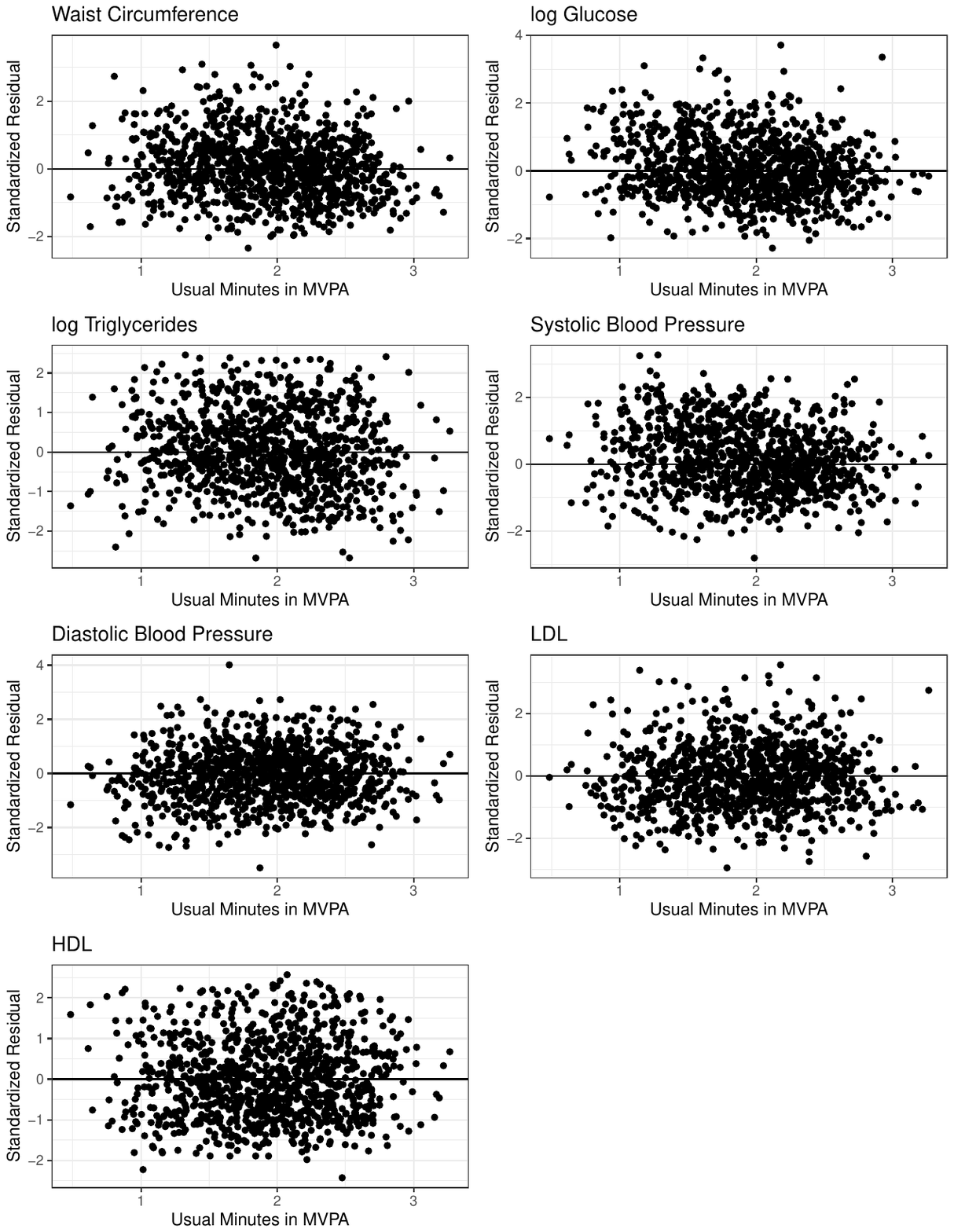}
\caption{Standardized residual plots for RFM using MEM-corrected MVPA. Y=0 line is bolded for reference.}
\label{fig:residplot_completecase}
\end{figure}

\begin{figure}[h]
\centering
\includegraphics[width=0.8\textwidth]{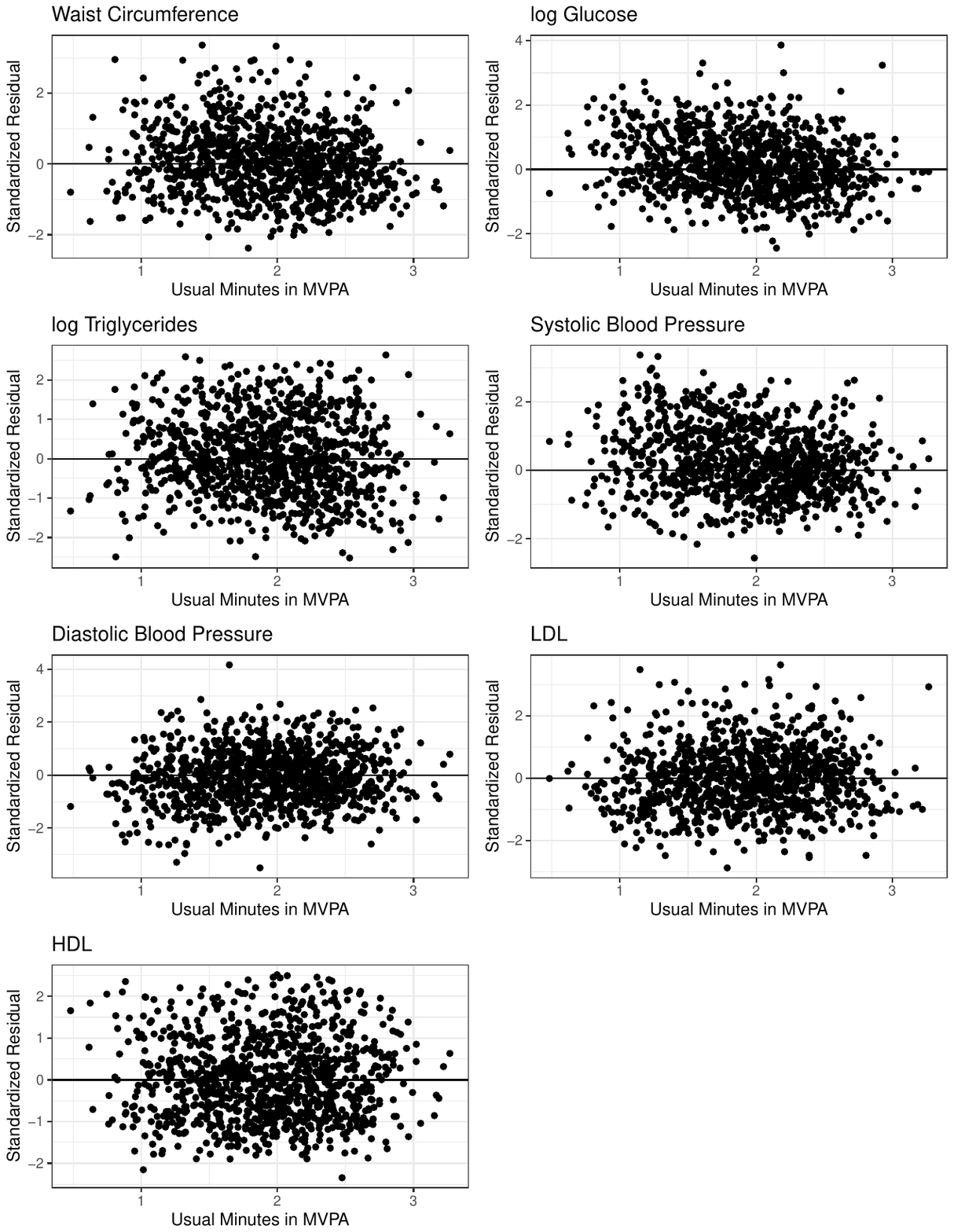}
\caption{Standardized residual plots for RFM using observed MVPA (no MEM). Y=0 line is bolded for reference.}
\label{fig:residplot_noMEM}
\end{figure}

\clearpage
\subsection*{Full conditional distributions for the RFM model parameters}

\begin{align}
	P(\zeta_i=h) | \cdot &\propto f({\bf Y}_i|\zeta_i,\theta) f(\zeta_i|{\bf p}) \propto p_h N({\bf \lambda}_h + {\bf m(t}_i,\boldsymbol{\gamma}),\Sigma_{m,h}) \\
	{\bf p}|\cdot &\propto f(\zeta_i|{\bf p}) f({\bf p}) = \text{Cat}(H,{\bf p}) \text{Dirichlet}(a) \sim \text{Dirichlet}(a+\tilde{n}) \\
	&\text{where} \\
	\tilde{n} &= (n_1,...,n_H) \\
	n_h &= \sum_{i=1}^{n} I(\zeta_i = h) \\
\boldsymbol{\lambda}_h |\cdot &\propto \prod_{i=1}^{n} f({\bf Y}_i|t_i,\zeta_i=h,\boldsymbol{\lambda}_h,\boldsymbol{\gamma},\Sigma_{m,h}) f(\boldsymbol{\lambda}_h) \sim N(m_{\lambda_h},V_{\lambda_h}), \forall h \\
&\text{where} \\
& V_{\lambda_h} = \left( V_0^{-1} + n_h \Sigma_{m,h}^{-1} \right) ^{-1} \\
& m_{\lambda_h} = V_{\lambda_h} \left( V_0^{-1} m_0 +  \Sigma_{m,h}^{-1} ({\bf Y^{(h)}-m(t}^{(h)},\boldsymbol{\gamma})) \right) \\
& {\bf Y}^{(h)} \text{ is response matrix for MetS risk factors for all individuals in component h.} \\
& {\bf t}^{(h)} = \{t_i\}_{\zeta_i = h}\\
\Sigma_{m,h}|\cdot &\propto f({\bf Y}_i|t_i,\zeta_i=h,\boldsymbol{\lambda}_h,\boldsymbol{\gamma},\Sigma_{m,h}) f(\Sigma_{m,h}) \sim \text{Inverse-Wishart}(d_0+n_h, D_0 + D) \\
D &= ({\bf Y}^{(h)} - \boldsymbol{\lambda}_h - {\bf m(t}^{(h)},\boldsymbol{\gamma}))' ({\bf Y}^{(h)} - \boldsymbol{\lambda}_h - {\bf m(t}^{(h)},\boldsymbol{\gamma})) \\
\boldsymbol{\gamma}|\cdot &\propto \prod_{i=1}^{n} f({\bf Y}_i|t_i,\zeta,\boldsymbol{\lambda}_h,\boldsymbol{\gamma},\Sigma_{m,h}) f(\boldsymbol{\gamma}) \\
&\propto  \prod_{i=1}^{n} e^{\frac{1}{2} ({\bf Y}_i - \boldsymbol{\lambda}_{h(i)} - {\bf m}(t_i,\boldsymbol{\gamma}))' \Sigma_{m,h(i)}^{-1} ({\bf Y}_i - \boldsymbol{\lambda}_{h(i)} - {\bf m}(t_i,\boldsymbol{\gamma}))} e^{\frac{1}{2} (\boldsymbol{\gamma} - m_{\gamma})'V_{\gamma}^{-1}(\gamma - m_{\gamma})} \\
&\text{where } h(i) \text{ is an indicator which component $Y_i$ belongs to}
\end{align}

\end{document}